\pdfoutput=1
\documentclass{JINST}
\usepackage{multirow}

\newcommand{\be}{\begin{equation}}
\newcommand{\ee}{\end{equation}}
\newcommand{\bea}{\begin{eqnarray}}
\newcommand{\eea}{\end{eqnarray}}

\newcommand{\libi}{$^{209}$Bi$^{80+}$}

\graphicspath{{figures/}}

\title{Detection system for forward emitted photons at the Experimental Storage Ring at GSI}

\author{V.~Hannen$^a$\thanks{Corresponding author.}, D.~Anielski$^a$, C.~Geppert$^{b,c}$, R.~J\"ohren$^a$, T.~K\"uhl$^c$, M.~Lochmann$^{b,c}$, R.~L\'opez-Coto$^a$, W.~N\"ortersh\"auser$^{b,d}$,  H.-W.~Ortjohann$^a$, R.~S\'anchez$^c$, J.~Vollbrecht$^a$, C.~Weinheimer$^a$ and D.F.A.~Winters$^c$\\
\llap{$^a$}Institut f\"ur Kernphysik, Westf\"alische Wilhelms-Universit\"at M\"unster, Wilhelm-Klemm-Str. 9, 48149 M\"unster, Germany\\
\llap{$^b$}Institut f\"ur Kernchemie, Johannes Gutenberg Universit\"at Mainz, Fritz Strassmann Weg 2, 55128 Mainz, Germany\\
\llap{$^c$}GSI Helmholtzzentrum f\"ur Schwerionenforschung GmbH, Planckstra{\ss}e 1, 64291 Darmstadt, Germany\\
\llap{$^d$}Institut f\"ur Kernphysik, Technische Universit\"at Darmstadt, Schlossgartenstr. 9, 64289 Darmstadt, Germany\\

  E-mail: \email{hannen@uni-muenster.de}
}

\abstract{
A single photon counting system has been developed for efficient detection of forward emitted fluorescence photons at the Experimental Storage Ring (ESR) at GSI. The system employs a movable parabolic mirror with a central slit that can be positioned around the ion beam and a selected low noise photomultiplier for detection of the collected photons. Compared to the previously used system of mirror segments installed inside the ESR the collection efficiency for forward-emitted photons is improved by more than a factor of 5. No adverse effects on the stored ion beam have been observed during operation besides a small drop in the ion current of about 5\% during movement of the mirror into the beam position. The new detection system has been used in the LIBELLE experiment at ESR and enabled for the first time the detection of the ground-state hyperfine M1 transition in lithium-like bismuth (\libi) in a  laser-spectroscopy measurement.
}


\begin{document}
\section{Introduction}
\begin{figure}[h]
\centering
\includegraphics[width=0.9\textwidth]{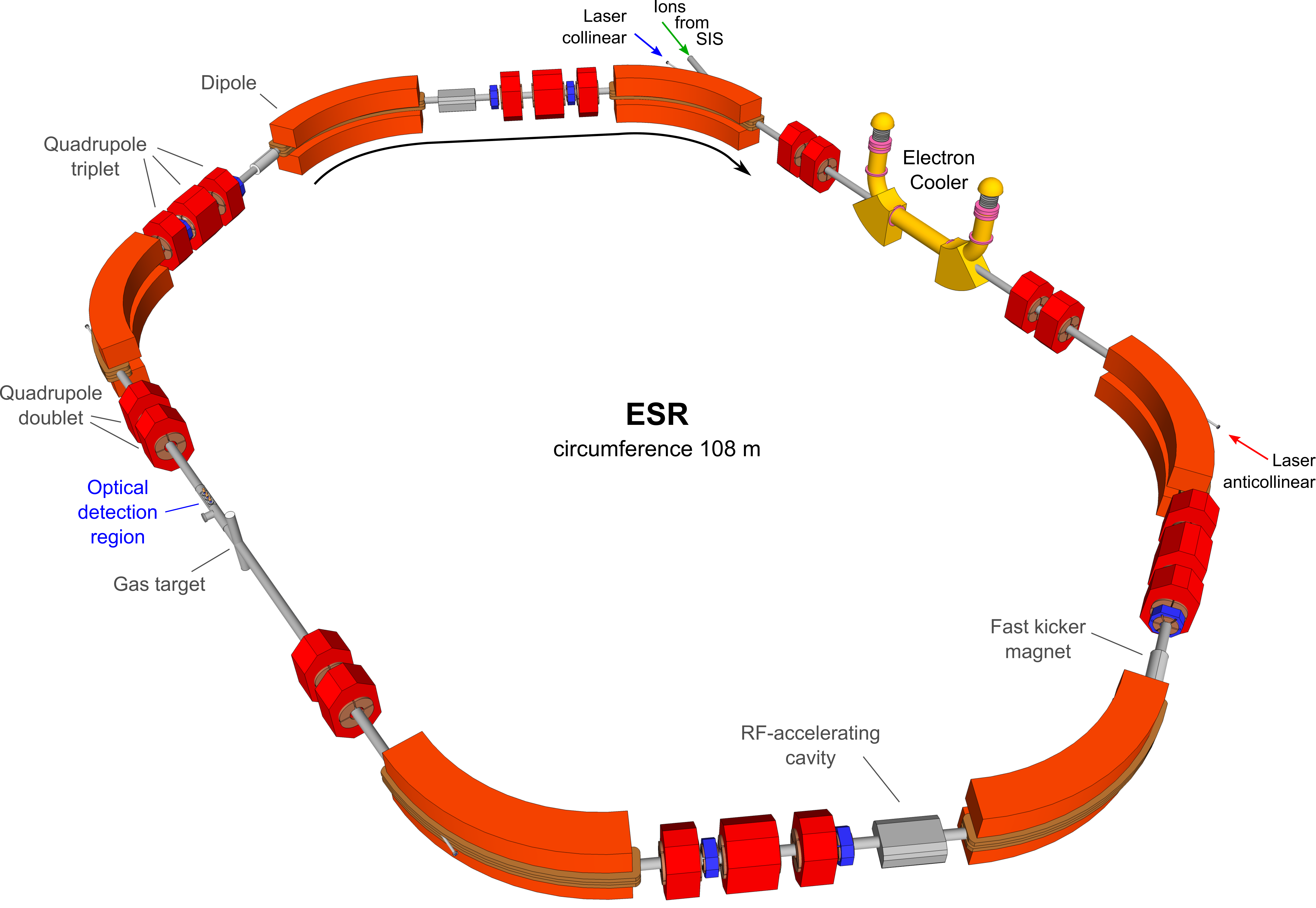}
\caption{Schematic view of the Experimental Storage Ring (ESR) at GSI with the optical detection region in the left straight section and the entrance windows for collinear or anti-collinear laser excitation of the ions inside the electron cooler of the ring on the right-hand side.}
\label{fig::esr}
\end{figure}

Knowing the hyperfine structure (HFS) transition of hydrogen- and lithium-like heavy ions of the same isotope allows one to test QED in extremely strong electromagnetic fields~\cite{sha01}. Bismuth is an especially important candidate for these studies as the wavelength of the HFS transition in both systems is accessible to laser spectroscopy experiments.\\
While the HFS transition energy of $^{209}$Bi$^{82+}$ has been determined in previous laser spectroscopy experiments at the Experimental Storage Ring (ESR) at GSI in Darmstadt~\cite{kla94}, the search for the HFS transition in \libi\ has, for a long time, been unsuccessful~\cite{win99,nor11}. Its detection is hampered by the long transition wavelength ($\lambda \approx 1555$~nm~\cite{sha00,vol09}) and a low signal rate due to the long lifetime of the HFS state of $\tau = 82$~ms~\cite{sha98}. 
The first problem can be overcome by storing the ions at high velocity inside a storage ring, where the wavelength of forward emitted fluorescence photons is Doppler shifted into the visible regime (e.g. for $\beta = 0.71$ to $\lambda \approx 640$~nm) and therefore becomes detectable with conventional photomultipliers. 
On the other hand, the low signal rate of the HFS transition in \libi\ becomes then even more problematic as the life-time of the excited ions in the lab-system increases due to relativistic time dilation (for $\beta = 0.71$ we have $\tau_{\rm lab} = 116$~ms). 
This is partially compensated by the concentration of the emitted radiation in a forward light cone, which reduces the solid angle into which light is emitted. It is therefore of special importance to have a detection system that collects forward-emitted fluorescence photons with a high efficiency.\\
Figure~\ref{fig::esr} shows an overview of the ESR with the optical detection region
and, on the opposite side of the ring, the entrance windows for collinear or anti-collinear laser excitation of the circulating ions inside the electron cooler section. The optical detection region contains a section with 10 conical mirror segments installed inside the beam pipe~\cite{see99}, that have been used in previous experiments to direct emitted photons through three viewports onto PMTs mounted on top of the beamline. However, as detailed in section~\ref{sec::seg_mirror}, this system mostly collects sideways emitted photons and is therefore not the right tool for the measurement of the HFS transition in \libi. This fact prompted the development of an in-beam mirror system that is specialized on collecting photons emitted under small angles with respect to the beam direction and that can be installed in addition to the existing system.\\
This article provides details of the newly developed detection system, while the data analysis and physics results of the LIBELLE experiment (GSI experiment number E083) will be explained in a forthcoming article by Lochmann et al.~\cite{Loc12}. Section~\ref{sec::mc} details Monte Carlo simulations that have been performed to devise an optimal design for the new system. Its mechanical construction is layed down in section~\ref{sec::parabolic_cad} and the performance in the actual experiment is described in section~\ref{sec::performance}.
\section{Monte Carlo simulations}
\label{sec::mc}
In order to investigate the reasons why previous experiments failed to detect the HFS transition in \libi with the existing mirror configuration and to optimize the design of the new in-beam detection system, optical tracking simulations have been performed~\cite{ani10} with a software package developed using the GEANT4 framework from CERN~\cite{geant4}. 
In the simulations we have assumed a total number of $2\cdot10^5$ excited ions to be stored in the ESR at a velocity of $\beta = 0.71$. The event generator of the simulation software produces fluorescence photons emitted within the 11~m long straight section of the ESR beampipe before the mirror section.
In the real experiment, however, photons are emitted along the complete circumference of the storage ring due to the long lifetime of the HFS state. This means that regardless of the system used, we can always detect only a fraction of all transitions.
The photons are emitted isotropically in the co-moving coordinate system of the ion and experience a Lorentz boost during the transformation to the laboratory system. Given the angle $\Theta$ under which an emitted photon is observed in the laboratory system relative to its direction of motion, the ratio of the corresponding solid angles in the co-moving system $\Delta \Omega'$ and in the laboratory system $\Delta \Omega$ is given by
\begin{equation}
  \frac{\Delta \Omega'}{\Delta \Omega} = \frac{1-\beta^2}{(1 - \beta\cos\Theta)^2}
\end{equation}
The wavelengths of the emitted photons are Doppler shifted according to 
\begin{equation}
  \lambda = \lambda' \cdot \gamma (1-\beta\cos\Theta) \; ,
\end{equation}
where $\lambda$ is the wavelength in the laboratory system and $\lambda'$ is the wavelength in the co-moving system.\\
Photon tracks that make it onto the photocathodes of one of the simulated PMTs are detected with a probability given by the wavelength dependent quantum efficiency (QE) of the detectors. Both in the existing mirror section and for the new movable mirror system we use selected PMTs of type R1017 produced by Hamamatsu~\cite{r1017} for detection of the fluorescence photons. These devices have a large active area with 46~mm diameter, a low dark-count rate and comparably high quantum efficiency of more than 10\% for photons with wavelengths between 500~nm and 600~nm that decreases to 1\% at 850~nm.
\subsection{Segmented mirror section}
\label{sec::seg_mirror}
\begin{figure}
\centering
\includegraphics[width=\textwidth]{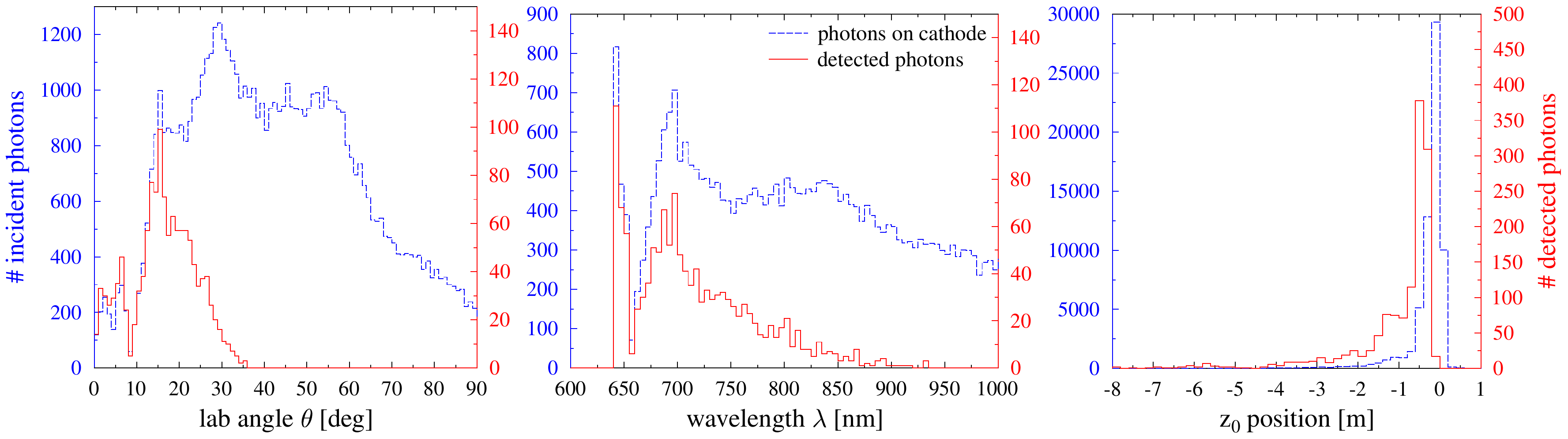}
\caption{Simulation results for the segmented mirror section. The dashed blue curves represent the distributions of emission angles, wavelengths and starting distances of those fluorescence photons
that are guided onto the photocathode of the PMT on window 3. The solid red curves are the corresponding distributions of photons that are actually detected, given the quantum efficiency of the PMT used.}
\label{fig::seg_results}
\end{figure}

The GEANT4 model of the segmented mirror section used in previous experiments~\cite{see99} 
\begin{figure}[b]
\centering
\includegraphics[width=0.48\textwidth]{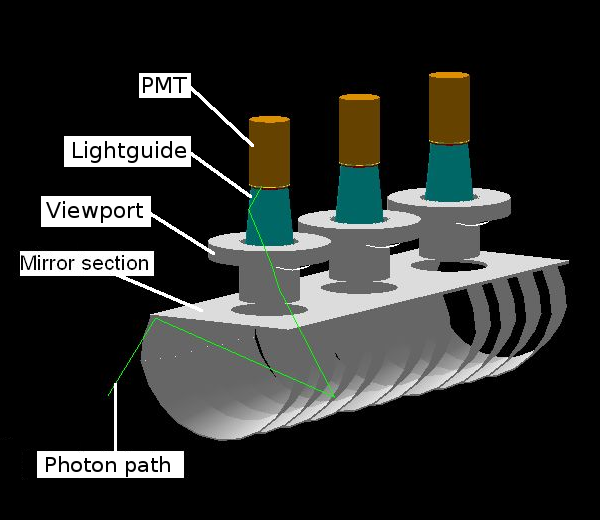}\hspace{2mm}
\includegraphics[width=0.48\textwidth]{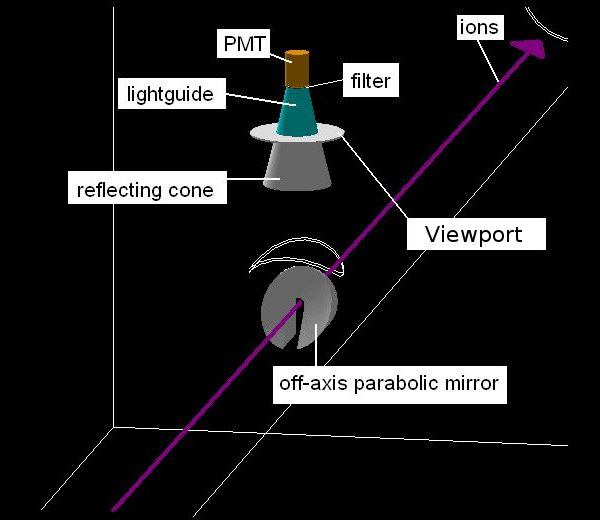}
\caption{Left: GEANT4 representation of the segmented mirror section. Right: Model of the parabolic mirror system. The beam pipe is omitted for clarity.}
\label{fig::seg_mirror}
\end{figure}
is shown on figure~\ref{fig::seg_mirror}, left side.
Fluorescence photons are reflected by one of 10 conical mirror segments and can exit the mirror system through one of three optical viewports on top of the mirror section. Outside the vacuum, photons are transported in lightguides towards the PMTs. To remove background photons at short wavelengths a long pass filter discarding wavelengths below 590~nm is inserted between lightguides and PMTs. 
Figure~\ref{fig::seg_results} shows results of the simulations for the third (downstream) window of the mirror section, which collected the
highest number of photons in the simulations. 
Only 3\% of the collected photons exhibit small emission angles in the range below $10^\circ$ and accordingly short wavelengths near 640~nm for forward emission. 
Higher angles that are more efficiently collected by the system make up the largest fraction of the detected photons, but are cut off around $30^\circ$ due to the lower QE of the PMT at the corresponding higher wavelengths. 
It also becomes obvious that most of the detected photons are emitted within the last 2~m before the detector. 
The sharp drops visible in the angular (left) and wavelength (middle) distributions are caused by transitions between the mirror segments, where certain light paths are reaching the edge of an individual mirror.
\subsection{Movable parabolic mirror}
\label{sec::parabolic_mc}
The GEANT4 representation of the newly developed parabolic mirror system is shown on the right side in figure~\ref{fig::seg_mirror}.
In this case, fluorescence photons are collected by an off-axis parabolic mirror with a central slit that allows the ion beam to pass the setup. Photons are directed towards a viewport window that is equipped on the vacuum side with an additional conical mirror made from a specially coated highly reflective aluminum~\cite{miro2} to increase the solid angle imaged onto the PMT. Outside the vacuum, photons are transported by a lightguide to a selected R1017 PMT. 
Like in the segmented mirror section, a 590~nm optical long pass filter is inserted between the lightguide and the PMT to cut off short-wavelength background photons. \\
The outer diameter of the parabolic mirror has been chosen to be 150~mm. There are two considerations that guided this decision. Firstly, one has to be able to retract the mirror far enough out of the beam position during injection of ions into the ring to not interfere with the uncooled ion cloud. Secondly, the experimental situation at the ESR made it necessary to mount the mirror on a horizontal lever arm that needs to be able to suspend the weight of the mirror and still move smoothly in and out of the central beam position.\\
The width of the central slit of the mirror was fixed by estimating the beam radius for the bismuth experiment from scraper measurements of the radius of a continous U$^{92+}$ beam which was stored in the ESR at a comparable $\beta$-value~\cite{Ste04}. 
In these measurements the beam radius was determined by moving a scraper into the ion beam and taking the difference between the position where the ion beam is absorbed completely and the position where a certain number of ions (i.e. a certain beam current) survives after the scraper has been moved to this position.
As the ions in the bismuth experiment are compressed into two bunches by an RF cavity in the storage ring, the density of ions in these bunches is about 10 times higher than for a continous beam with the same current. Extrapolating the uranium beam radius as measured with the scrapers to an ion density comparable to the bunched \libi\ beam at 3~mA, we obtain radii of 7.2~mm in the horizontal and 5.5~mm in the vertical direction. About two times the estimated horizontal beam radius was considered a safe choice for the radius of the central opening of the mirror and the slit width was therefore set to 30~mm.\\
The remaining free parameter of the mirror design is the focal length $f'$ towards the PMT
\begin{figure}
\centering
\includegraphics[width=0.52\textwidth]{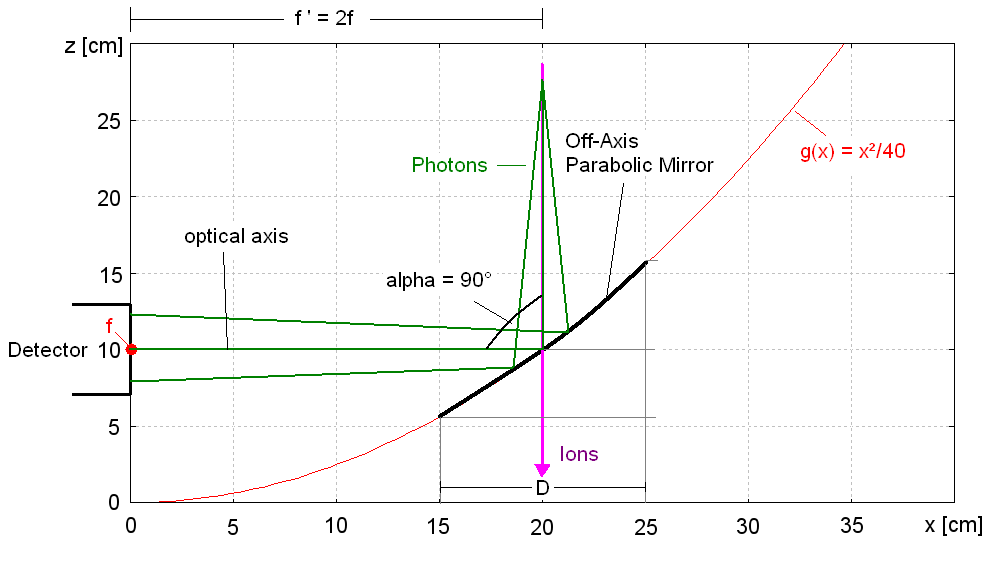}\hspace{2mm}
\includegraphics[width=0.4\textwidth]{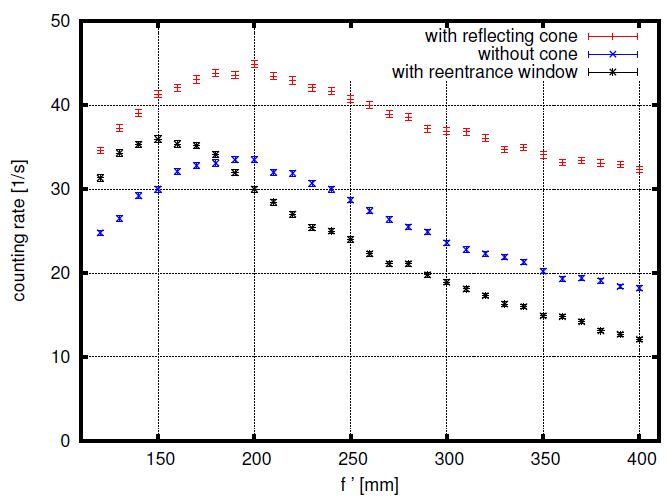}
\caption{Left: geometrical situation at the parabolic mirror setup. Right: simulated count rates as a function of the focal length $f'$.}
\label{fig::focal_length}
\end{figure}
(see figure~\ref{fig::focal_length}, left). This parameter has been optimized with respect to the achievable count rate for three different geometries: 1) the parabolic mirror setup as described above, 2) the same setup but without the reflecting mirror cone and 3) a setup without the aluminum cone, but with a special viewport, where the viewport window is smaller compared to the standard situation (72~mm free diameter instead of 89~mm) but is located 110~mm closer to the beam axis.  
Figure~\ref{fig::focal_length}, right, shows the result of this optimization. As expected the overall count rate is lower when omitting the reflecting aluminum cone. Also the special viewport geometry does not deliver the same detection rate as obtained with geometry 1). The most suitable focal length for the setup with aluminum cone is found at $f' = 200$~mm
(for this optimization we have assumed a PMT with standard quantum efficiency and zero reflectivity of the beam-pipe material).\\
Taking into account the possibility of using selected PMTs with quantum efficiencies up to 1.4 times the standard QE and assuming a reflectivity of 25\% for the beam-pipe material, we obtain simulated distributions for the emission angle, wavelength and distance from the  
\begin{figure}[h]
\centering
\includegraphics[width=\textwidth]{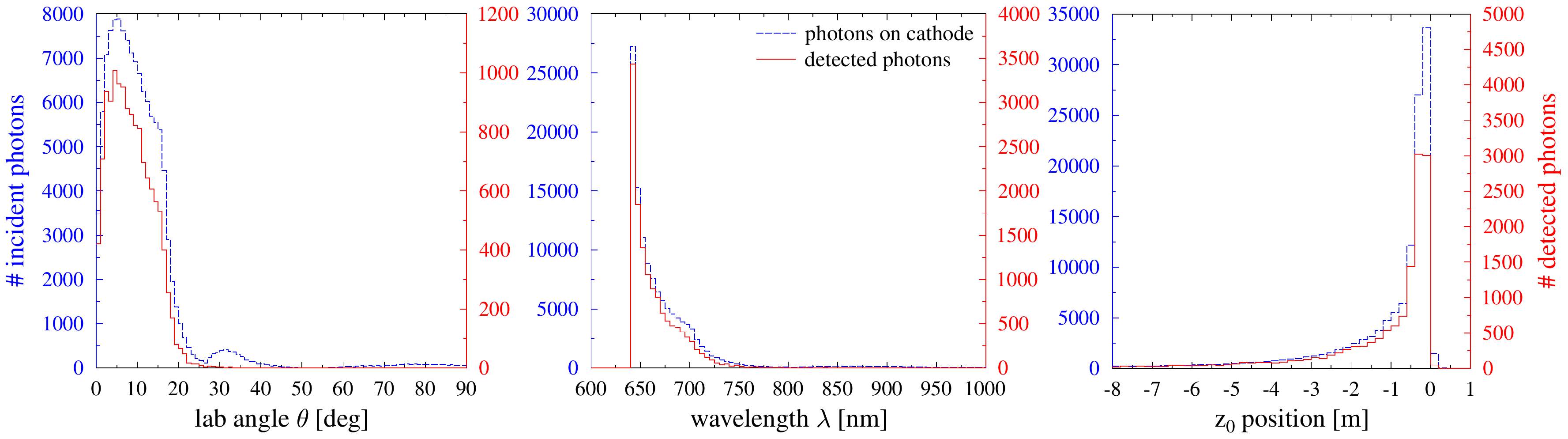}
\caption{Simulation results for the movable parabolic mirror. The dashed blue curves represent the distributions of emission angles, wavelengths and starting distances of those fluorescence photons that are guided onto the photocathode of the PMT. The red curves are the corresponding distributions of photons that are actually detected, given the quantum efficiency of the PMT used.}
\label{fig::para_results}
\end{figure}
mirror center for photons detected with the new setup as displayed in figure~\ref{fig::para_results}.
As intended, the parabolic mirror mainly collects forward emitted photons with a peak at about $5^\circ$ emission angle. The wavelengths of these photons are mostly smaller than 730~nm, where the quantum efficiency of a standard R1017 PMT is still 5\%. Detected photons are for the most part emitted within 2~m distance of the parabolic mirror center.\\
Table~\ref{tab::sim_results} provides a comparision between simulated single-photon count rates for the segmented 
\begin{table}[b]
\centering
\begin{tabular}{cccc}
\hline
                         &  signal rate [s$^{-1}$] & background rate [s$^{-1}$] & measurement time [s] \\
\hline
Segmented mirror section & $15.4\pm0.5$    & $626\pm7$        & $47.6\pm0.9$ \\
Parabolic mirror         & $86.0\pm0.6$    & $453\pm9$        & $1.1\pm0.01$  \\
\hline
\end{tabular}
\caption{Simulated signal and background count rates for the segmented mirror section and the new movable parabolic mirror setup. In the simulation we assumed a selected type R1017 PMT with a quantum efficiency 1.4 times higher than for a standard PMT of the same type. We also assumed a reflectivity of 25\% for the stainless steel material of the beam pipe. Background rates were simulated using a simple model of excited rest-gas molecules distributed homogeneously in the beam pipe. The quoted measurement times are the times required per wavelength step for a detection of the signal with $3 \sigma$ significance. The quoted errors are purely statistical.}
\label{tab::sim_results}
\end{table}
mirror section and for the new parabolic mirror setup.
The background rates were simulated using a simple model of excited rest gas molecules distributed homogeneously in the beam pipe. In this model, background photons are produced with uniformly distributed wavelengths between 590~nm (below which no photons reach the detectors due to the filters placed in front of the PMTs) and 940~nm (above which the QE of the PMTs is negligible).
The overall amount of background photons was adjusted to reproduce background count rates observed during previous experimental attempts to detect the \libi HFS transition with the segmented mirror section. It was also taken into account that the data acquisition system is only active during those intervals when the ions, which are cycling the ESR in two bunches, are passing the detector. The resulting duty cycle of the data acquisition was assumed to be 20\%, which corresponds to two bunches with a length of approximately 10~m.\\
Concluding we can state that the simulations of the two detection systems predict an improved signal rate observed with the new parabolic mirror setup that is a factor 5.6 times higher than for the segmented mirror section while at the same time reducing the experimental background by about 30\%. The required measurement time per wavelength step (see table~\ref{tab::sim_results}) for a detection of the fluorescence signal with $3 \sigma$ significance is thus a factor 43 lower for the new setup. 
\section{Mechanical design}
\label{sec::parabolic_cad} 
\begin{figure}[h]
\centering
\includegraphics[width=0.6\textwidth]{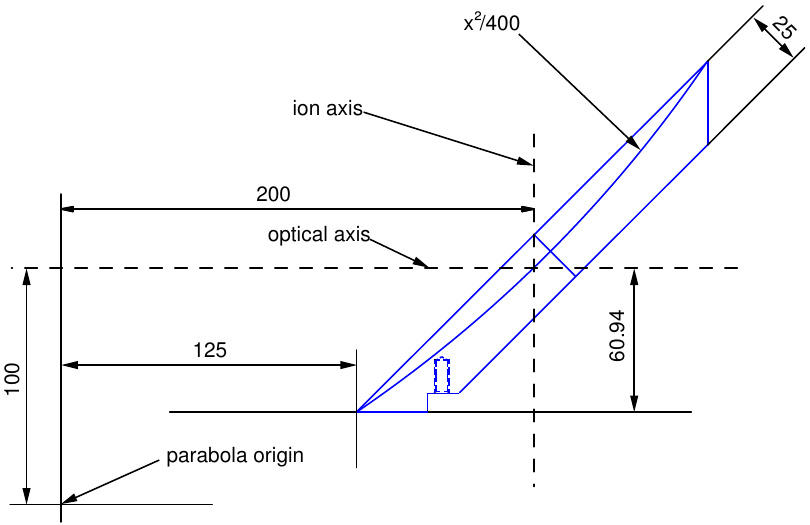}\hspace{5mm}
\raisebox{10mm}{\includegraphics[width=0.35\textwidth]{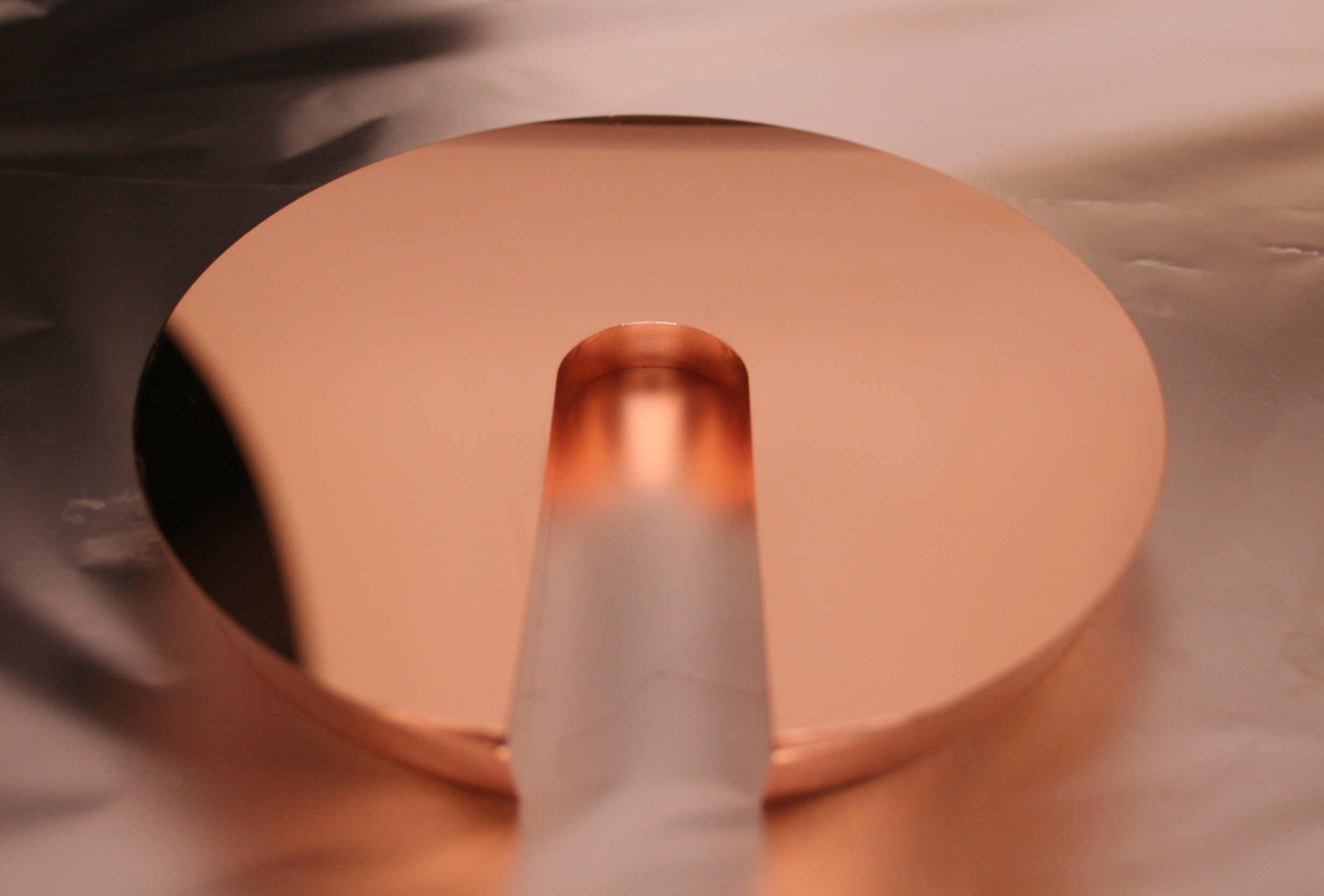}}
\caption{Left: mirror dimensions in mm, right: polished mirror made from OFHC copper material.}
\label{fig::mirror_cad}
\end{figure}

\begin{figure}[h]
\centering
\includegraphics[width=\textwidth]{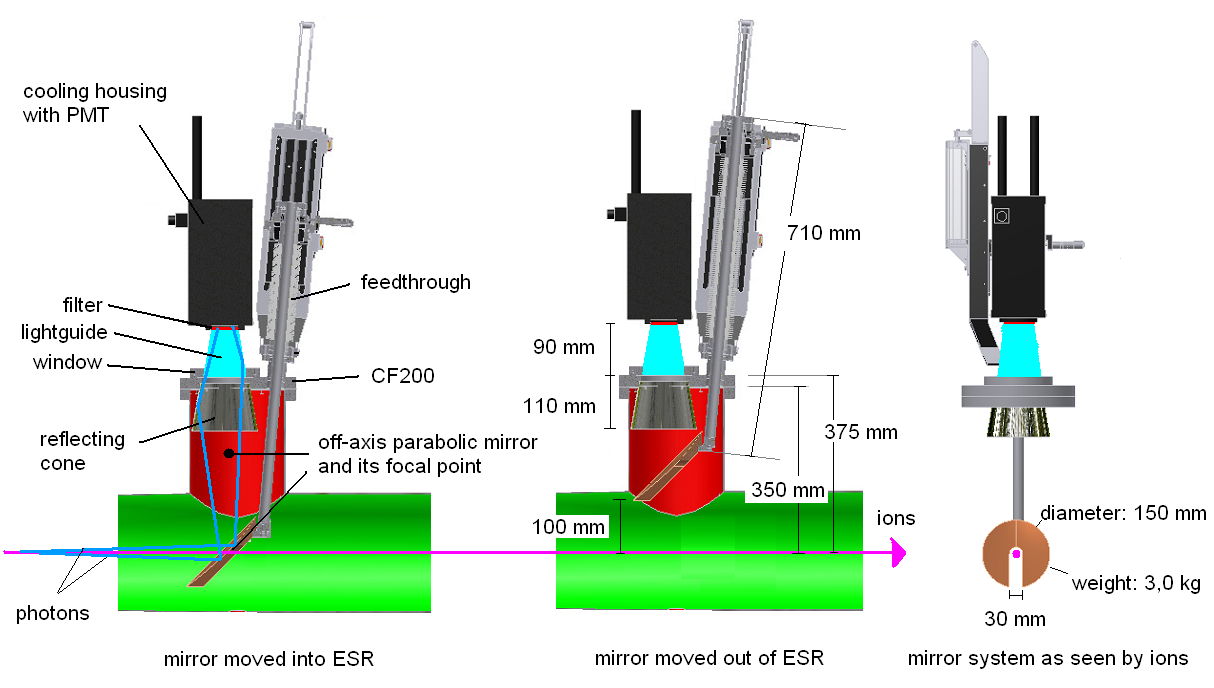}
\caption{CAD model of the parabolic mirror system installation at the ESR beamline. It should be noted that the mirror system is mounted horizontally.}
\label{fig::setup_cad}
\end{figure}

\begin{figure}[b]
\centering
\includegraphics[width=\textwidth]{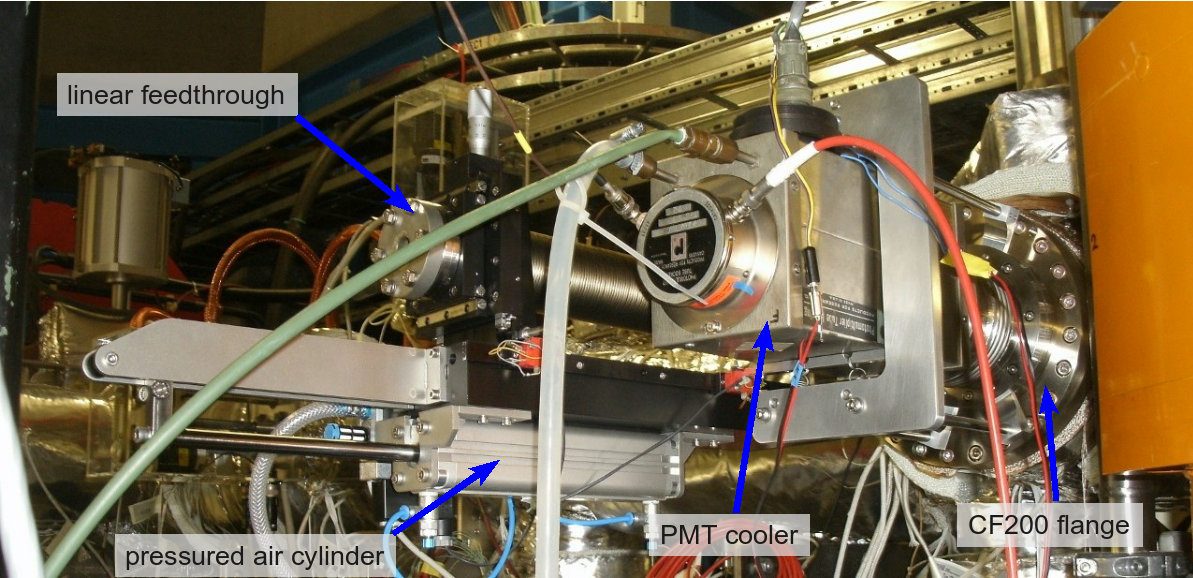}
\caption{Parabolic mirror setup mounted at the ESR.}
\label{fig::mirror_esr}
\end{figure}

Figure~\ref{fig::mirror_cad} displays the geometry that results from the simulations together with a photograph of the finished mirror. For the mirror material we have 
chosen oxygen-free (OFHC) copper, which is compatible with the ultra-high vacuum requirements of the ESR and has a high reflectivity ($>95\%$) for wavelengths above 620~nm. Some tests have been performed with a gold plating to protect the copper material against oxidation. Like copper, gold has a high reflectivity at the wavelengths of interest. Unfortunately the gold layers deposited by evaporation onto the copper samples did not withstand bake-out at $350^\circ$C, which is one of the requirements for use at the ESR.
After heating in a vacuum oven, most of the gold had diffused into the copper material, leaving a surface with a strongly reduced reflectivity. Also a layer of nickel deposited between the copper and the gold layer, did not reliably stop the diffusion process and also resulted in a strongly reduced reflectivity after bake-out. In contrast, pure copper did not show any degradation of its reflectivity after the heating process. 
Additionally, aging effects due to oxidation of polished copper have been shown by Lester and Saito~\cite{les77} to have little effect on the reflectivity at 630~nm. Nevertheless, whenever possible the mirror has been stored in an argon atmosphere or under vacuum.\\
Figure~\ref{fig::setup_cad} shows an overview of the experimental setup at the ESR. The mirror is moved in and out of the beam 
position by an air pressure driven, linear feedthrough with 150~mm stroke. Photons reflected from the mirror are collected on the vacuum side by the aluminum cone (made from MIRO2, a specially coated material from Alanod~\cite{miro2}) and are transmitted through a viewport window made from kodial glass with 89~mm free diameter. On the atmosphere side, the photons are then guided by a plexiglas lightguide to a combination of a Schott OG590 long pass filter~\cite{sch08} and a Balzers CALFLEX~X heat protection filter~\cite{calflex}. These serve as an optical bandpass transmitting wavelengths from 590~nm to 725~nm and thus removing background photons outside our region of interest.
Photons that pass the filters are detected by a selected Hamamatsu R1017 PMT~\cite{r1017}. To reduce thermal noise, the PMT is cooled to approximately $-20^\circ$C using a PR-TE104RF cooler from Products for Research Inc.. The linear feedthrough is slightly tilted to alleviate spatial constraints on the CF200 flange onto which the components are mounted. Figure~\ref{fig::mirror_esr} shows a photograph of the parabolic mirror 
setup at the ESR. Because the whole assembly is mounted horizontally, a double-walled stainless-steel rod has been manufactured to hold the approximately 3~kg copper mirror without bending down too much. To allow for a smooth movement of the mirror in and out of its beam position, an additional counter weight on a pulley has been installed (not shown on the photo). This was used to balance the torque caused by the heavy mirror on the linear feedthrough and, at the same time, to counteract the pull of the ESR vacuum.\\
For readout of the PMT we first amplified the signal by a factor 100 using a CAEN N979 fast amplifier followed by a CAEN N843 constant-fraction discriminator to produce a logical output signal that could then be processed by a counter or time-to-digital converter (TDC) unit.
During the LIBELLE experiment, the logical signals were processed by a GSI developed VUPROM (VME Universal Processing Module) unit~\cite{Hof06} which has been configured as a multihit TDC with additional scaler channels.
\section{Experimental Results}
\label{sec::performance}
\begin{figure}[h]
\centering
\includegraphics[width=0.6\textwidth]{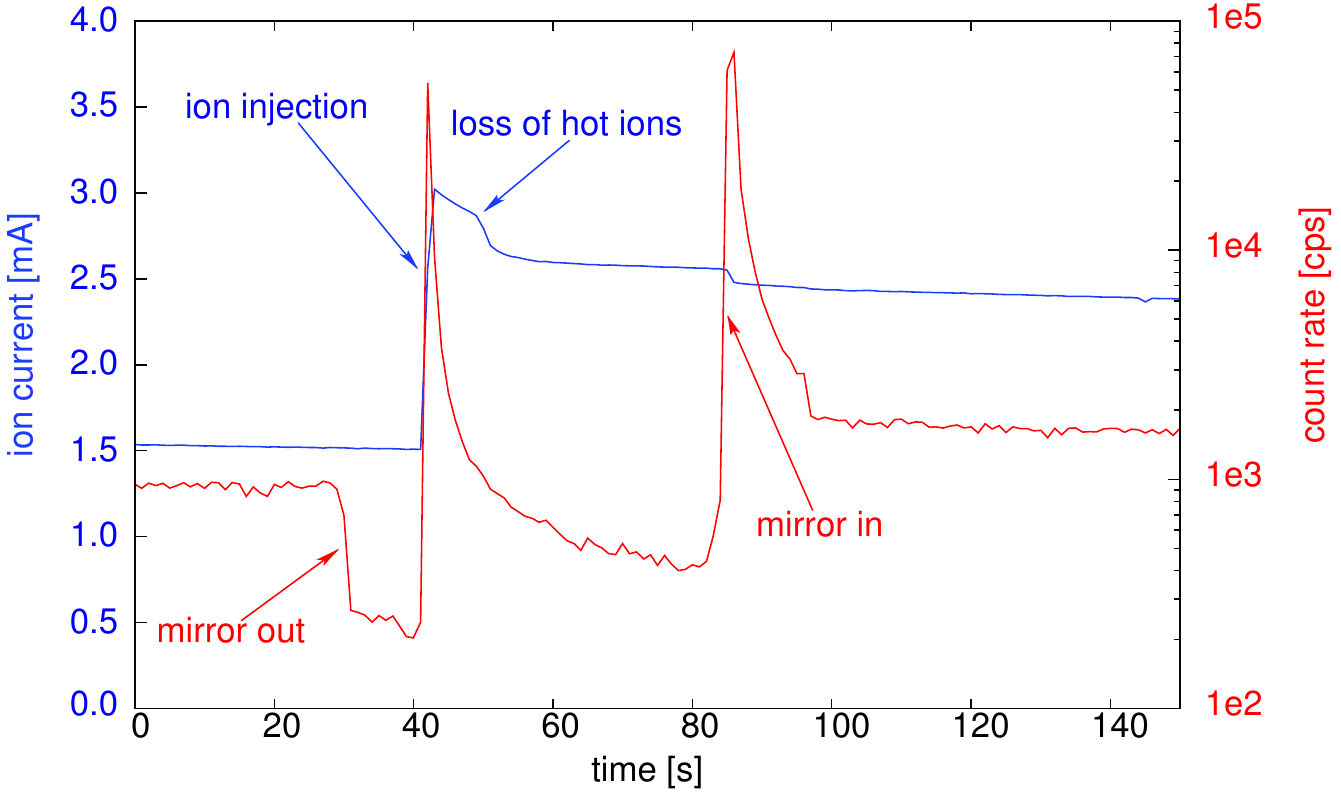}
\caption{Time evolution of background count rates and corresponding ion current in the ESR during movement of the mirror and injection of new \libi\ ions. The ion current is represented on a linear scale (left axis), whereas the PMT count rate is plotted on a logarithmic scale (right axis).}
\label{fig::current}
\end{figure}

Figure~\ref{fig::current} displays the time evolution of count rates obtained with the parabolic mirror setup 
during the LIBELLE Experiment~\cite{Loc12} together with the corresponding ion current in the ESR.
The plot illustrates the different phases of a measurement cycle, where the mirror is, at a low ion current, withdrawn from the beamline (at about 30~s), upon which new \libi\ ions are injected into the ring (at 40~s). Even with the mirror retracted, the uncooled ion cloud causes a substantial background count rate in the detector, which decreases over the next 40~s while the ion beam is cooled by the electron cooler of the ring. After cool-down of the beam, the mirror is inserted into the beamline again (at about 80 s) and a sharp rise in the detector count rate is observed, caused by the impact of ions from the tails of the beam profile on the copper mirror. This beam halo is only slowly re-populated, e.g.\ by intra-beam scattering, but still contributes an important part of the experimental background during the measurements (see section~\ref{sec::background}).
Once the high background from ion-impact events has levelled and decreases only slowly with beam current (at roughly 110~s), the laser spectroscopy measurements can resume.
\subsection{Background rates}
\label{sec::background}
\begin{figure}[b]
\centering
\includegraphics[width=0.48\textwidth]{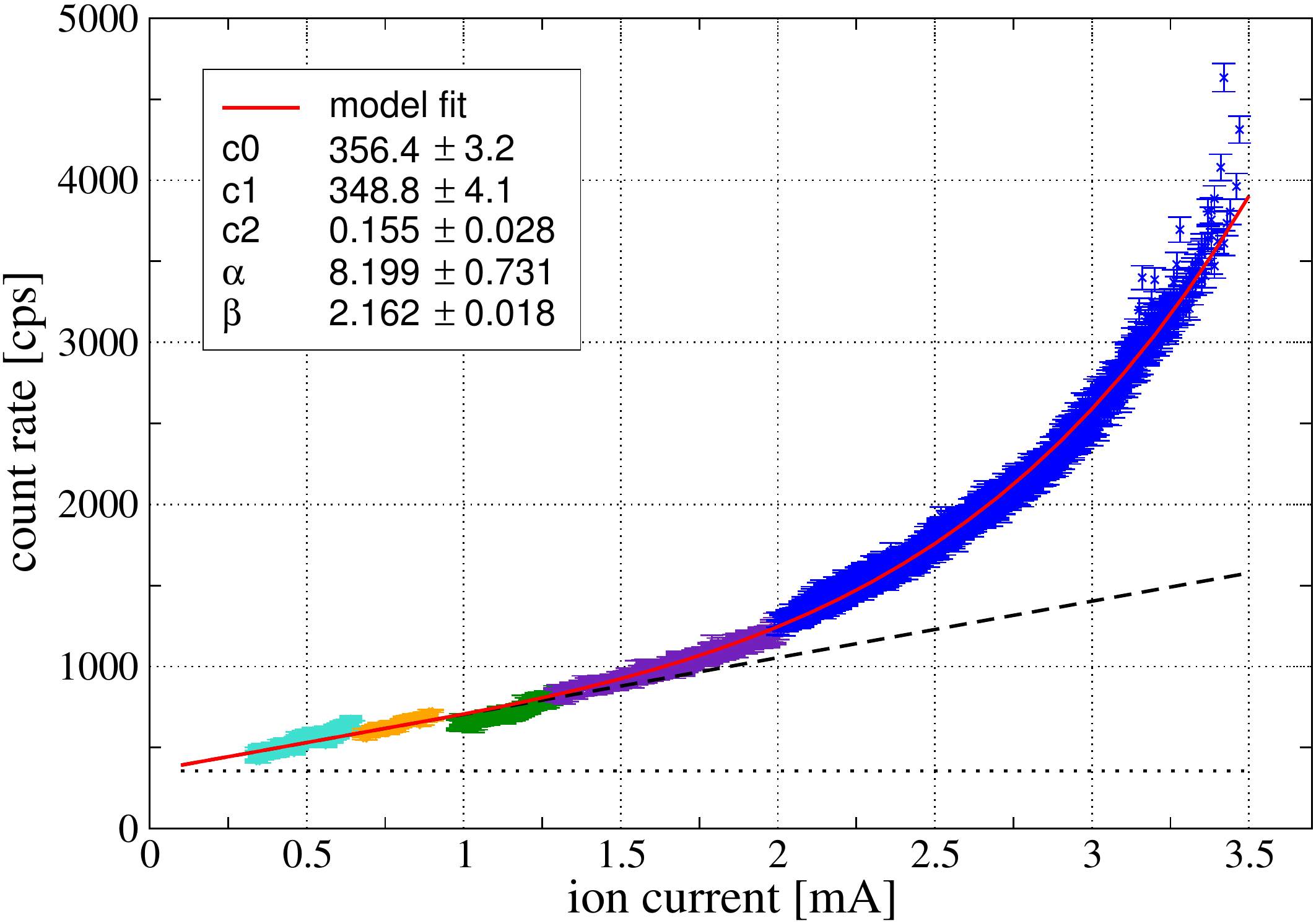}
\includegraphics[width=0.48\textwidth]{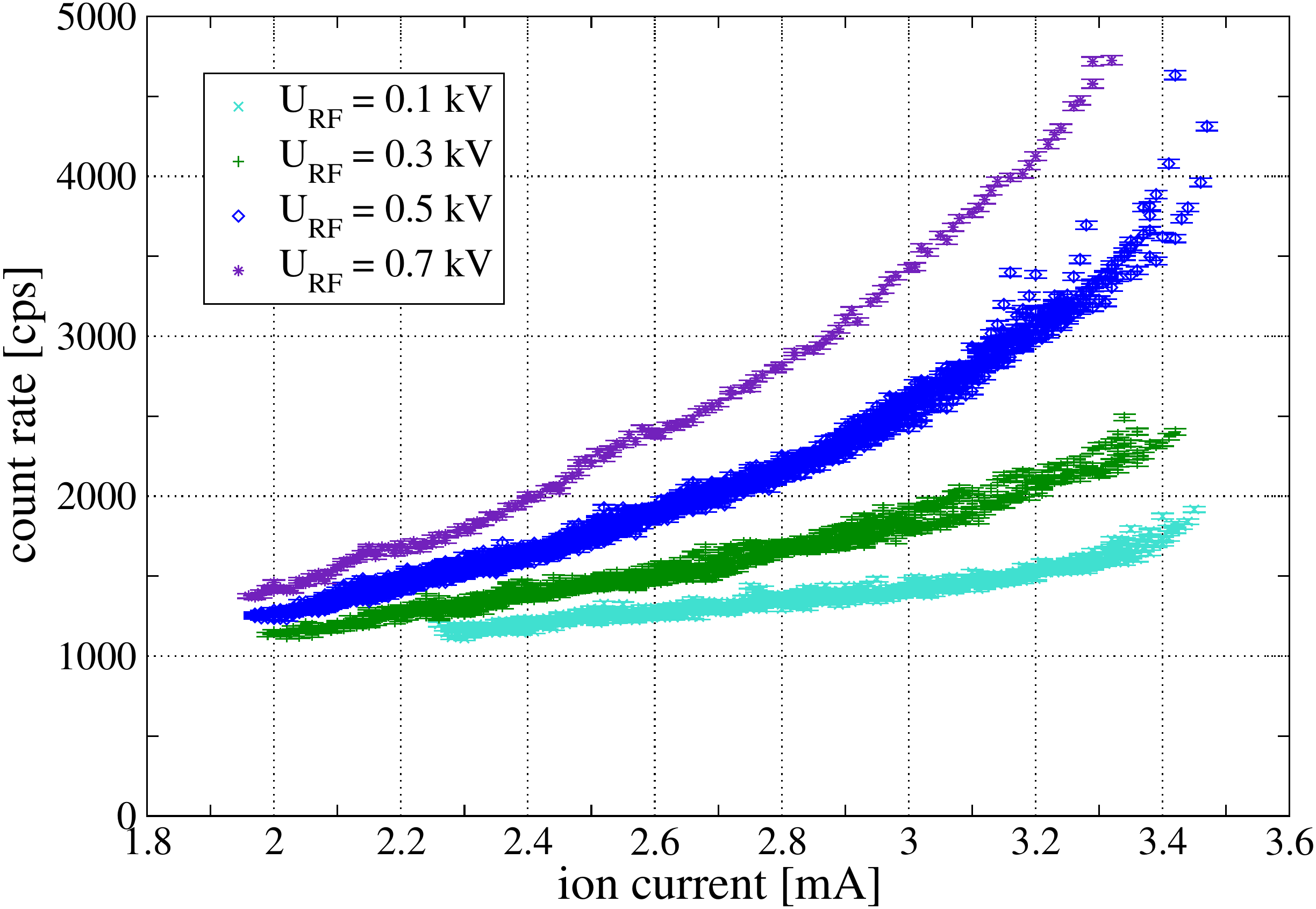}
\caption{Left: Measured PMT count rates as a function of ion current in the ESR. The red line represents the model fit, while the dotted and dashed lines display the constant and linear contributions to the fit, respectively. To cover the range of ion currents displayed, data from several measurement phases had to be collected as indicated by the different colors of the data points. 
Right: Dependence of the background rate on the amplitude of the RF signal used to divide the two ion bunches.}
\label{fig::background}
\end{figure}

The count rate, which is observed during the active phases of the experiment is mainly determined by background processes, even when the laser wavelength hits the HFS resonance of the \libi\ ions. 
There are four sources contributing to this background: The dark count rate of the PMT $D_{\rm PMT}$ (which for the cooled device is $< 100$~cps), photons originating from light leaks in the storage ring $D_{\rm ll}$ (e.g. due to not perfectly sealed windows in the beamline), fluorescence photons from excitation of rest gas molecules by the ion beam $D_{\rm rg}(I)$ and, finally, background photons produced by the impact of ions from the beam halo onto the mirror material $D_{\rm h}(I)$. 
The overall background rate is then given by 
\begin{equation}
 D(I) = D_{\rm PMT} + D_{\rm ll} + D_{\rm rg}(I) + D_{\rm h}(I) \; ,
\end{equation}
where the last two terms depend on the beam current $I$. 
While the background from excited residual gas ions $D_{\rm rg}(I)$ is assumed to be proportional to the beam current, the background from ions hitting the mirror has a more complex dependence on the current.
In a simple model, we assume the beam profile to be described by a radially symmetric Gaussian distribution
$f(r,\phi)$ and can then calculate the number of ions outside a certain radius $r_0$ using
\begin{equation}
 \int_0^{2\pi} \int_{r_0}^\infty f(r,\phi) r \, {\rm d}r \, {\rm d}\phi 
   = \int_0^{2\pi} \int_{r_0}^\infty A \exp\left(-\frac{r^2}{2\sigma^2}\right) r\, {\rm d}r \, {\rm d}\phi 
   = 2\pi A \sigma^2 \exp\left(-\frac{r_0^2}{2\sigma^2}\right) \; ,
\end{equation}
with $A$ being the amplitude of the gaussian, and the rms radius $\sigma$ assumed to be the same for the horizontal and vertical beam axis. 
According to Steck et al.~\cite{Ste04}, the dependence of the rms radius $\sigma$ on the beam current can be described by a power law, i.e. $\sigma = \alpha \, I^\beta$, such that we can use the following model to fit our observed background:
\begin{equation}
 D(I) = c_0 + c_1 \cdot I + c_2 \cdot \alpha^2 \, I^{2\beta} \exp\left(-\frac{r_0^2}{2\alpha^2 \, I^{2\beta}}\right) \; .
\end{equation}
Figure~\ref{fig::background}, left plot, shows background rates as a function of beam current 
collected from several measurement phases together with the model fit. The error bars of the experimental data points were scaled by the square root of the reduced $\chi^2$ value in order to account for the imperfect matching of the included data sets, which cannot be reproduced by the model fit.
The combined background rate of PMT dark counts $D_{\rm PMT}$ and light leaks $D_{\rm ll}$ is given by the constant term and is found to be $(356\pm 3)$~cps. This can be compared to a prior measurement of the rate where there was no ion beam in the ring that yielded a rate of about 220~cps. The higher count rate extrapolated from the measurements with beam could be due to a slightly higher temperature of the photomultiplier, as the power supply of the PMT cooler failed at the beginning of the beam time and had to be exchanged with a different model.\\
During the \libi\ experiment, ions orbiting in the ESR were divided into two bunches by applying an RF high-voltage signal at the second harmonic of the free revolution frequency of the ions to a cavity in the storage ring. The amplitude of this signal influences the shape of the ion bunches and, therefore, also the ion density in the bunches. At a lower RF voltage, the ion bunches become larger in the longitudinal direction, while the beam radius decreases due to the 
lower intrabeam ion scattering rate resulting from the lower ion density per unit length. 
This is visible in figure~\ref{fig::background}, right plot, where the measured background rates are shown for different values of the bunching amplitude $U_{\rm RF}$.
At the lowest voltage $U_{\rm RF} = 0.1$~kV the rate stays linear with current up to high beam currents. The nonlinear contribution caused by the impact of ions from the outer regions of the beam profile, only becomes visible above 3.2~mA.
\subsection{Fluorescence detection}
\begin{figure}[h]
\centering
\includegraphics[width=0.7\textwidth]{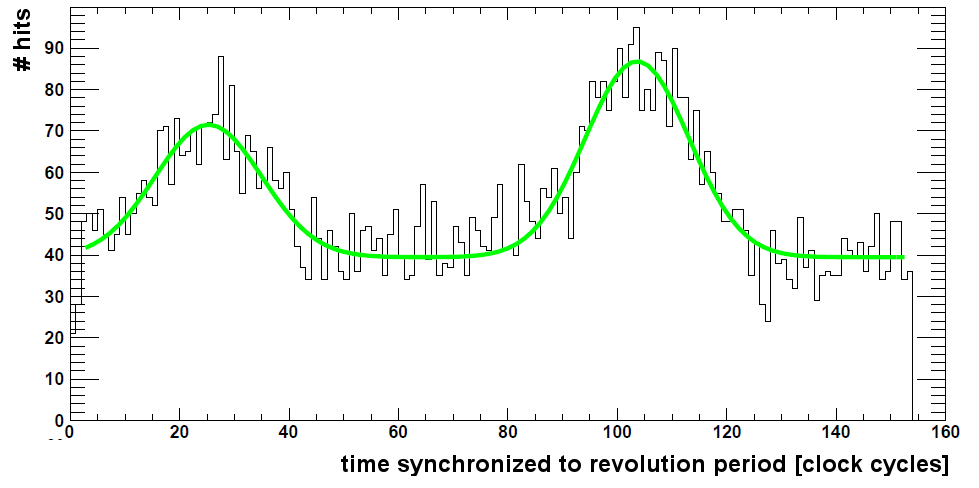}
\caption{Timing spectrum of PMT signals accumulated during a 5.94~s interval with the laser wavelength being on resonance with the HFS transition in Li-like bismuth.
The time is given in cycles of the 300~MHz clock of the TDC used for readout.
The green line corresponds to a fit consisting of a constant background term and two gaussians representing signals from the two ion bunches. The left peak is caused by the reference bunch, while the right peak includes photons of the laser-excited signal bunch.}
\label{fig::stacked}
\end{figure}

For the spectroscopy measurements, only one of the bunches (the signal bunch) was excited by a pulsed laser synchronized to the $\approx 2$~MHz revolution frequency of the ions. The other one (the reference bunch) was used for background substraction.  
The amplified and discriminated PMT signals were recorded by a multi-hit TDC operated in common-stop mode. The common-stop signal was provided by a scaled-down version (1/200) of the RF bunching frequency. In the analysis, TDC hits from individual cycles of the ion bunches in the ESR were stacked on top of each other in a timing histogram covering the time period required for a single revolution. 
In this way, all counts belonging to a certain wavelength step of the tunable laser used for excitation of the ions were accumulated in a single spectrum. \\
Figure~\ref{fig::stacked} displays such a timing histogram for the case where the laser wavelength was on the resonance 
frequency of the HFS transition in \libi .
The spectrum is fitted by a constant background with two gaussians on top comprising the counts from the reference bunch (left peak in the plot) and the signal bunch (right peak). 
From the difference of the peak areas we calculate the number of fluorescence photons from the HFS transition to be $344\pm 78$. Taking into account the accumulated measurement time of 5.94~s, we obtain a signal rate of $(58\pm 13)$~cps. The average ion current during this wavelength step was 2.2~mA, while the simulated results reported in table~\ref{tab::sim_results} were obtained assuming a current of 3~mA. To compare the experimental result to the simulation we therefore have to scale the count rate by the ratio of currents to obtain $(79\pm 18)$~cps, which compares well to the rate of 86~cps expected from simulation (see table~\ref{tab::sim_results}). \\
To compare the observed background rate to the Monte Carlo results, where a 20\% duty cycle of the data acquisition was assumed (see section~\ref{sec::parabolic_mc}), we select two 15 bin wide regions around the peak centers in figure~\ref{fig::stacked} that correspond to 20\% of the revolution period. In these regions we find altogether $(2058\pm 64)$ background counts. If we scale this number with the ratio of actual and simulated current and normalize to the measurement time, we arrive at an experimental background rate of  $(472 \pm 15)$~cps that is close to the simulated rate of 453~cps, which was obtained from a simplified background model (see table~\ref{tab::sim_results}).
For the latter comparision it should be noted that in the experiment we had an additional NIR blocking filter~\cite{calflex}, that removed some background at long wavelengths.
\section{Summary}
We have constructed a novel detection system for fluorescence photons emitted by highly charged ions stored in the ESR. The system uses a parabolic mirror with a central slit that can be positioned around the ion beam to collect photons emitted under very small angles with respect to the beam direction and focus those photons onto a PMT placed outside the vacuum. The detector was successfully used in the LIBELLE experiment and enabled for the first time the detection of the HFS transition in Li-like bismuth in a laser spectroscopy measurement~\cite{Loc12,nor13}. The signal and background rates determined from the experimental data are in good agreement with results obtained from a GEANT4 simulation during the design phase of the instrument. 
\acknowledgments
We would like to thank Markus Steck from GSI for fruitful discussions regarding the mirror requirements and his support during the LIBELLE beamtime.
This work has been supported by BMBF under contract numbers 06MS9152I and 05P12RDFA4, and from the Helmholtz Association under contract VH-NG-148. M.L. acknowledges financial support from HIC for FAIR.

{}
\end{document}